\documentclass[11pt]{amsart}
\usepackage[dvips]{graphics,color}
\usepackage{amsmath}
\usepackage{amsfonts,amssymb,latexsym}
\usepackage{amscd}
\usepackage{mathrsfs}
\usepackage{amsthm}
\theoremstyle{plain}
\newtheorem{theorem}{Theorem}

\theoremstyle{remark}
\newtheorem*{example}{Example}

\newtheorem*{remarks}{Remarks}

\usepackage{graphics}
\numberwithin{equation}{section}
\def\Bbb{\mathbb}

\def\logo{\raisebox{-10.5\p@}{\hb@xt@85\p@{\includegraphics{gft.eps}\hfil}}}

\def\un{1\kern-3pt \rm I}
\def\ptoday{{\ifcase\month 
\or January, \or February, \or March, \or April,\or May, 
\or June, \or July, \or August, \or September, \or October, 
\or November, \or December,\fi\ \number \year}}

\input epsf
\usepackage[dvips]{graphicx}

\def\dj{\hbox{d\kern-0.347em \vrule width 0.3em height 1.252ex depth
-1.21ex \kern 0.051em}}

\begin{document}


\title[On Renormalization in Finite Temperature Field Theory]
      {\sl Microlocal Analysis and Renormalization \\[3mm]
       in Finite Temperature Field Theory}
        
\author{Daniel H.T. Franco}
\address{Centro de Estudos de F\'\i sica Te\'orica, Setor de F\'\i sica--Matem\'atica\\
         Rua Rio Grande do Norte 1053/302, Funcion\'arios \\
         Belo Horizonte, Minas Gerais, Brasil, CEP:30130-131.}
\email{dhtf@terra.com.br}

\author{Jos\'e L. Acebal}
\address{Centro Federal de Educa\c c\~ao Tecnol\'ogica de Minas Gerais\\
         Avenida Amazonas 7675, Nova Gameleira\\
         Belo Horizonte, Minas Gerais, Brasil, CEP: 30.510-000.}
\email{acebal@dppg.cefetmg.br}

\date{August 03, 2006}
\begin{abstract}
We reassess the problem of renormalization in finite
tem\-pe\-ra\-tu\-re field theory (FTFT). A new point of view elucidates the relation
between the ultraviolet divergences for $T=0$ and $T \not= 0$ theories and
makes clear the reason why the ultraviolet behavior keeps unaffected
when we consider the FTFT version associated to a given quantum field
theory (QFT). The strength of the derivation one lies on the H\"ormander's criterion
for the existence of products of distributions in terms of the wavefront
sets of the respective distributions. The approach allows us to regard the FTFT both
imaginary and real time formalism at once in a unified way in the contour ordered formalism.
\end{abstract}

\maketitle


\,\,\,{\hfill{\it To Prof. Olivier Piguet}}

\section{Introduction}
As it occurs in QFT, the FTFT also exhibit ultraviolet divergences.
The problem of how to make sense out of the physical meaning behind the divergences
in a mathematically proper way was satisfactorily solved by the known renormalization
procedure. There are some well established prescriptions currently used in QFT to
attribute meaning to the initially divergent terms of the 
perturbation series associated to the quantities of interest. The latter 
can however be defined only up to certain renormalization ambiguities which, in 
principle, can be determined from physical reasonings. 
In facing the distinctions between the FTFT and QFT 
propagators, some questions take place. Once the divergences are related to certain 
ill-defined products of distributions, the FTFT propagator might imply
changes in the conditions for the existence of the products and introduce
a temperature-dependent renormalization problem. The ambiguities of the 
renormalization procedure associated to the physical parameters 
could then exhibit qualitative changes due to the temperature-dependence. 
Further, it could also change the asymptotic divergent behavior and 
consequently the amount of arbitrariness involved. The FTFT propagator being
separable into temperature-dependent and -independent pieces, causes the mixing
of the divergences and temperature-dependent terms in crossing products in 
the higher order terms of the perturbation expansion. Depending on
the renormalization procedure adopted some of those facts can become not clear.
On physical grounds, one cannot expect that the differences would
have fundamental consequences to the UV behavior because it arises from the 
short distance limit which is unaffected by the temperature once the thermal 
part of the propagator has support on the mass shell and decays rapidly with 
growing momentum because of the Bose-Einstein (or Fermi-Dirac) distribution 
function. This question is not really new and has been investigated by many
authors using various techniques, each one putting emphasis on different aspects
of the problem. Let us mention for instance the TFD proof~\cite{MaOjUm}, 
the RFT method~\cite{NiSe}, the BPHZ momentum space subtraction 
procedure~\cite{LandsWeert,MaKo} besides the framework of axiomatic quantum 
field theories at finite temperature~\cite{Ste}. More recently, it has been 
given by C. Kopper {\em et al.}~\cite{KoMuRe} a rigorous proof of the 
renormalizability of the massive $\varphi^4_4$ theory at finite
temperature based in the framework of Wilson's flow equations, to all orders of 
the loop expansion.

In this article, we take the opportunity to shed a new light on the series of studies
by approaching the problem from a central aspect of the renormalization which lies
on the lack of  definition of the distributional product in some particular context
present in  the perturbation series. The analysis is done under the light of a systematic
use of the ideas and notions of the distribution theory. Microlocal methods of
distributions in $x$-space are used in order to characterize the singular spectrum
in terms of the wavefront set of the propagators and to determine a sufficient
condition for the existence of such products~\cite{Hor1}. The asymptotic behavior of
products of distributions near the singular points is evaluated by calculating the scaling 
degree and the singular order of distributions~\cite{Ste1}-\cite{HW}
which govern the amount arbitrariness present in the renormalization procedure.
The whole apparatus provide us with all the information we need in order to formulate
the renormalization as the well posed mathematical problem of the extension of products
of distributions to coincident points~\cite{Stora,BF}. One shows that the
divergences found in FTFT are, in fact, of the same nature
as those ones in QFT. At each order, the problem of the extension in FTFT is shown
to reduce to the analogous one of the ordinary QFT. As a consequence, it is proved that
the amount of arbitrariness in the renormalization procedure, as well as the type of the
ambiguities remain the same when passing from a given QFT to the associated FTFT version.
More important, our analysis allows to investigating the issue, as much as possible, in
a model independent way and free from the technical difficulties of thermal loop
calculations common to the various conventional approaches both in ITFs and RTFs.
Moreover, it allows to adopting the generalized unified framework of the contour ordered
formalism (COF) considering at a time both ITF and RTF.

The outline of the article is as follows. We begin in Section 2 by describing some basics
on the microlocal analysis of singularities where the wavefront set of a distribution is 
introduced together with a sufficient condition for the existence of products of 
distributions based on its wavefront set. In Section 3, we reproduce a derivation
of the FTFT free two-point function, and we prove the required wavefront set properties
of this two-point function, in order to be able to insert it in the
renormalization scheme. This section also includes a discussion of some
examples. The extent of the analysis of these examples is to indicate
where extra singular terms may come into the picture -- over and those
above appearing in usual QFT at $T=0$. Section 4 contains the final considerations.

\section{Microlocal Study of Singularities}
\label{MSS}
The UV divergences are a QFT inherent problem, because the fields, as well as
its correlation functions, having distributional character are defined on a
continuous space-time. The perturbation expansions in QFT are made of the product
of such distributions. However, products of distributions with overlapping
singularities are in general not well-defined. Hence, it becomes convenient to
shed some light on the problem of finding the conditions under which one has or not a
well-defined product of distributions.
Among the distributional analysis techniques, the framework of the 
{\em microlocal analysis}~\cite{Hor1} is fairly suitable for the study of the UV 
divergences. The term microlocal analysis refers to a set of techniques of 
relatively recent origin which have turned out to be particularly useful in 
analyzing partial differential equations with variable coefficients, including 
those of particular interest to quantum field theory.
In what follows, we shall describe an analytical method which provides
sufficient conditions for the existence of the product of distributions based
on the concept of the {\em wavefront set} (WFS) of a distribution $f$, denoted 
by ${WF}(f)$. It is a refined description of the singularity spectrum.
More important, WFS not only describes the set where a distribution is
singular, but also localizes the frequencies that constitute these singularities.  
Similar notion was developed in some versions by Sato~\cite{Sa} and 
Iagolnitzer~\cite{Ia}. The present definition is due 
to H\"ormander \cite{Hor1} who has made use of this terminology due to an 
existing analogy between the ``propagation'' of singularities of distributions 
and the classical construction of propagating waves by Huyghens.

Let $f$ be a distribution on an open set $X \subset {\Bbb R}^d$; then the
{\em singular support} of $f$ is the complement of the largest relatively
open subset $X^1$ of $X$ whereon $f$ is {\em smooth}
($f|_{X^1} \in C_0^\infty$). A point $x_0$ is said to be a
{\em non-singular point} of a distribution $f$ if there exists a cutoff
function $\phi \in C_0^\infty(V)$, with support in some neighborhood $V$ of
$x_0$, such that the Fourier transform
\[
\widehat{f \phi}(k)=\int d^dx\,\,f(x)\phi(x)e^{ikx}\,\,,
\]
is of fast decrease for all directions $k \in {\Bbb R}^d$. By a fast decrease
in the $k$ direction of $\widehat{f}(k)$, one must understand that there is a
constant $C_N$, for all $(N=1,2,3\dots)$, such that $(1+|k|)^N|\widehat{f}(k)|\leq C_N$ 
remains bounded. In particular, if $x_0$ is a singular point of the distribution $f$, 
and $\phi \in C_0^\infty(V)$ is such that $\phi(x_0)\not= 0$; then $\phi f$ is 
also of compact support and singular in $x_0$. In this case, can still occur some
directions in $k$-space over which $\widehat{\phi f}$ is asymptotically bounded. A 
direction $k$ for which the Fourier transform $\widehat{f}(k)$ of 
$f(x)\in{\mathscr D}^\prime(V)$ shows to be of fast decrease is called to be a 
{\em regular direction} of $\widehat{f}(k)$. This suggests that we can single out 
singular directions as well as singular point, and for the establishment
of these concepts only the behavior of $f$ and of $\widehat{f}$ restricted to
an arbitrarily small neighborhood of the singular point $x_0$ is relevant. 

Let $f(x)$ be an arbitrary distribution not necessarily of 
compact support on an open set $X \subset {\Bbb R}^d$. Then, the set of all 
pairs composed first by the its singular points $x\in X$ and second by the 
associated nonzero singular directions $k$, 
\begin{equation}
{WF}(f)=\Bigl\{(x_0,k) \in X \times ({\Bbb R}^d\backslash 0)\left|\right. k 
\in \Sigma_x(f)\Bigr\}\,\,, 
\label{A.2}
\end{equation}
is called {\bf wavefront set} of $f$.  The $\Sigma_x(f)$ is defined to be the 
complement of the set of all $k \in {\Bbb R}^d\backslash 0$ with respect to
${\Bbb R}^d\backslash 0$, for which there is an open conic neighborhood $M$
of $k$ such that $\widehat{\phi f}$ is of fast decrease on $M$.
In short, to determine whether $(x_0,k)$ is in WFS of $f$ one must 
first to localize $f$ around $x_0$, to next obtain Fourier transform
$\widehat{f}$ and finally to look at the decay in the direction $k$.

\begin{example} A small ``point'' scatterer on ${\Bbb R}$.
\[
V(x)=\delta(x)\propto \int d^dx\,\,{\bf 1}e^{-\,ikx}\,\,,
\]
{\em i.e.}, $\widehat{V}={\bf 1}$ does not decay in any direction $k$:
$WF(\delta)=\{(0,k) \mid k \not= 0\}$ has singularities in all directions.  
\end{example}

\begin{remarks}
We now collect some basic properties of the WFSs:
\begin{enumerate}

\item The ${{WF}}(f)$ is conic in the sense that it remains invariant under the 
action of dilatations, {\em i.e.} when one multiplies the second variable by a 
positive scalar. If $(x,k) \in {{WF}}(f)$ then $(x,\lambda k)\in {{WF}}(f)$ 
for all $\lambda > 0$. \label{wfsdilat}

\item From the definition of the wavefront set, it follows that the projection onto 
the first coordinate $\pi_1({{WF}}(f)) \rightarrow x$, consists of those points 
that have no neighborhood whereon $u$ is a smooth function, and the projection 
onto the second coordinate $\pi_2({{WF}}(f)) \rightarrow \Sigma_x(f)$, is the 
cone around $k$ attached to a such point de\-no\-ting the set of high-frequency
directions responsible for the appearance of a singularity at this point.

\item The WFS of a smooth function is the empty set.\label{wfssmooth}

\item For all smooth function $\phi$ with compact suport 
${WF}(\phi f)\subset {WF}(f)$. \label{wfsfgsmooth}

\item For any partial linear differential operator $P$, with $C^\infty$ 
coefficients, one has \[{WF}(Pf)\subseteq {WF}(f)\,\,.\label{wfsop}
\]

\item If $f$ and $g$ are two distributions belonging to 
${\mathscr D}^\prime({\Bbb R}^d)$, with wavefront set ${WF}(f)$ and ${WF}(g)$, 
respectively; then the wavefront set of 
$(f+g) \in {\mathscr D}^\prime({\Bbb R}^d)$ is contained in 
${WF}(f)\cup{WF}(g)$. \label{wfssum}

\end{enumerate}
\label{WFSPROP}
\end{remarks} 

In the perturbation scheme of quantum field theories, one finds formal 
operations on distributions which can be in general not well-defined. 
In order to give precise statements on the existence of the product of these 
distributions, we appeal to a criterion based on the WFS of the distributional 
factors the so-called {\bf H\"ormander's Criterion}. Let $u$ and $v$ be 
distributions; if the WFS of $u$ and $v$ are such that the following direct sum
\begin{align}  
{WF}(u) \oplus {WF}(v) \overset{\text{def}}{=}
\Bigl\{(x,k_1+k_2) \mid (x,k_1)\in {WF}(u),(x,k_2)\in {WF}(v)\Bigr\}\,\,,
\label{HC}
\end{align}
does not contain any element of the form $(x,0)$, then the product $uv$ there 
exists and ${WF}(uv)\subset{WF}(u)\cup{WF}(v)\cup({WF}(u)\oplus{WF}(v))$.
Hence, the product of the distributions $u$ and $v$ is well-defined around
$x$, if $u$, or $v$, or both distributions are regular in $x$. Otherwise, if
$u$ and $v$ are singular in $x$, the product can still exist if the sum of the 
second components from ${WF}(u)$ and ${WF}(v)$ related to $x$ can be 
linearly combined with nonnegative coefficients to vanish only by a trivial 
manner.

\begin{example}
The distributions $u,v \in {\mathscr D}^\prime({\Bbb R})$, 
$u(x)=\frac{1}{x+i\epsilon}$ and $v(x)=\frac{1}{x-i\epsilon}$, with the
Heavyside distributions $\widehat{u}(k)=-2\pi i\theta(-k)$ and 
$\widehat{v}(k)=2\pi i\theta(k)$ as their Fourier transforms, have the
following WFSs:    
\begin{align*}
{WF}(u)=\Bigl\{(0,k) \mid k \in {\Bbb R}^-\backslash 0\Bigr\}\,\,,\quad
{WF}(v)=\Bigl\{(0,k) \mid k \in {\Bbb R}^+\backslash 0\Bigr\}\,\,.
\end{align*}
Thus, from the H\"or\-man\-der's Criterion one finds that there exist the powers 
of $u^n$ and $v^n$. However, the product between $u$ and $v$ do not match the 
criterion above and do not exist. This example clearly indicates that one 
can multiply distributions even if they have overlapping singularities, provided 
their WFSs are in favorable positions. Such an observation is significant 
because it makes clear that {\em the problem is not only where the support is, 
but in which directions the Fourier transform is not rapidly decreasing}!

\end{example}

\begin{example} The Feynman pro\-pa\-ga\-tor for massive scalar field 
\begin{equation}
\Delta_{\rm F}(x)\overset{\text{def}}{=}
\theta(x^0)\Delta_+(x;m^2)-
\theta(-x^0)\Delta_-(x;m^2)\,\,,
\label{fpropag}
\end{equation}
can have its WFS constitution studied from the WFS of the Wightman functions,  
\begin{align}
{WF}(\Delta_\pm)=&\Bigl\{((0,\bf{0});(\pm\lambda|{\bf k}|,\mp\lambda{\bf k}))\mid 
(\bf{k}\neq 0) \in {\Bbb R}^3, \lambda \in {\Bbb R}_+ \Bigr\} \nonumber \\[3mm]
&\cup \Bigl\{((|{\bf x}|,{\bf x});(\pm\lambda|{\bf k}|,\mp \lambda{\bf k}))
\mid {\bf x},(\bf{k}\neq 0) \in{\Bbb R}^3,\,\lambda \in {\Bbb R}_+ \Bigr\}\,\,,
\label{wfwight}\\[3mm]
&\cup \Bigl\{((-|{\bf x}|,{\bf x});(\pm\lambda|{\bf k}|,\mp \lambda{\bf k}))
\mid {\bf x},(\bf{k}\neq 0) \in{\Bbb R}^3,\,\lambda \in {\Bbb R}_+ \Bigr\}\,\,,
\nonumber
\end{align}
and from the WFS of $\theta(\pm t \mp t^\prime)\overset{\rm def}{=}\theta^\pm$,
\begin{align}
{WF}(\theta\pm)=\Bigl\{((0,{\bf x});(\pm \lambda k_0,\bf{0}))\mid {\bf x}
\in {\Bbb R}^3, k_0\in{\Bbb R}, \lambda \in {\Bbb R}_+ \Bigr\}\,\,.
\label{wftheta}
\end{align}
One can easily conclude that is not possible to form a non trivial linear 
combination with non-negative coefficients in order to produce a vanishing 
second component in the direct sum of the WFSs above.
So,
\begin{align}
(x,0)\not\in{WF}(\theta\pm)\oplus {WF}(\Delta_\pm)\,.
\label{dirsum}
\end{align}
Therefore, from the H\"{o}rmander's criterion, the Feynman propagator can be 
well-defined in terms of the product above and 
\begin{align}
{WF}(\theta^\pm\cdot\Delta_\pm)
\subset\,\,{WF}(\theta^\pm)\cup{WF}(\Delta_\pm)
\cup ({WF}(\theta^\pm)\oplus{WF}(\Delta_\pm))\,.
\label{superset}
\end{align}
However, in the powers $(\Delta_{\rm F})^n$ there exist products like 
$\Delta_+\Delta_-$ and from (\ref{wfwight}), one can see that 
$(x,0)\in{WF}(\Delta_+) \oplus {WF}(\Delta_-)$ and it occurs for the singular 
point $x=0$. In this sense, one must be careful when manipulating such products.
In fact, they are known to exist anywhere, except at $x=0$. Such an 
ill-definition, manifested as divergences, requires the treatment of the 
renormalization. Notice further that 
\begin{align}
(x,0)\not\in{WF} (\Delta_\pm)\oplus{WF}(\Delta_\pm).
\label{w2pm}
\end{align}
In particular, it can be used
\[
\Delta_\pm(x; m^2)=\frac{\pm i}{(2\pi)^3}\int d^4k_1\,\,\theta(\pm k_1^0)
\delta(k_1^2-m^2)\,e^{-\,ik_1x }\,\,,
\]
and $\widehat{\Delta}_\pm(k_1,k_2)=\pm i(2\pi)^{4}\delta(k_1+k_2)
\theta(\pm k_1^0) \delta(k_1^2-m^2)$ as a representation of the Fourier 
transform, to verify that the wavefront set of Feynman propagator has the 
following covariant form~\cite{Rad}:
\begin{align*}
WF(\Delta_{\rm F})
=&\Bigl\{(x_1,k_1);(x_2,k_2)\in ({\Bbb R}^{1,3}\times{\Bbb R}^{1,3}
\setminus 0)\mid
x_1 \not= x_2, (x_1-x_2)^2=0, k_1 \parallel (x_1-x_2), \\[3mm]
&\qquad \qquad \qquad k_1+k_2=0, k_1^2=0, k_1^0 > 0\,\,{\text{if}}\,\,x_1 \succ x_2
\,\,{\text{and}}\,\,k_1^0 < 0\,\,{\text{if}}\,\,x_1 \prec x_2 \Bigr\} \\[3mm]
& \cup \Bigl\{(x_1,k_1);(x_2,k_2)\in({\Bbb R}^{1,3}\times{\Bbb R}^{1,3}
\setminus 0)\mid 
x_1=x_2, k_1+k_2=0, k_1^2=0 \Bigr\}\,\,,
\end{align*}
where we have used the notation that $x_1 \succ x_2$ if $x_1-x_2$ is in the convex
hull of the forward lightcone and $x_2 \succ x_1$ if $x_1-x_2$ is in the convex
hull of the backward lightcone.
Notice that the condition $k_1^0 > 0\,\,{\text{if}}\,\,x_1 \succ x_2\,\,
{\text{and}}\,\,k_1^0 < 0\,\,{\text{if}}\,\,x_1 \prec x_2$ in 
$WF(\Delta_{\rm F})$ ensures the existence of products of Feynman propagators at 
all points away from diagonal, while these products do not satisfy the 
H\"ormander's criterion for multiplication of distributions over the points of 
the diagonal, since the sum of the second components of the WFS on the 
diagonal can add up to zero.
\label{WFFP}
\end{example}

\section{Renormalization of Distributions in FTFT}
In order to study the structure of the renormalization scheme in FTFT, we turn 
to the analysis of distributions and their products present in the 
perturbation series. The existence of such products are checked out via the H\"ormander's 
criterion based on its WFSs. Keeping in mind the renormalization procedure as an 
extension problem~\cite{Stora,BF} together to the its inherent arbitrariness governed by
scaling degree and singular order of distributions~\cite{Ste1}-\cite{BF}, the perturbation
expansion is further discussed. Without loss in generality, let us consider the
case of a single, scalar field $\phi(x)$ in FTFT associated to spinless 
particles with mass $m>0$,\footnote{The generalization of the present prescription to 
any field with arbitrary spin is straightforward.} whose propagator is given by: 
\begin{equation}
G^{c}(x,x^\prime)=
\theta_c(t-t^\prime)\langle\widehat{\phi}(x)\widehat{\phi}(x^\prime)\rangle +
\theta_c(t^\prime-t)\langle\widehat{\phi}(x^\prime)\widehat{\phi}(x)\rangle\,\,.
\label{propag1}
\end{equation}
The brakets $\langle \cdots \rangle$  stand for statistical average related to 
states of a complete orthogonal basis in Fock space. The index ``$c$'' 
accounts for the contour ordering in the complex time plane-t $(t=x_0+ix_4)$
whose the imaginary and real parts are interpreted to be the inverse temperature
and actual time respectively. For the contour ordering prescription given by 
$\theta_c(t-t^\prime)$, it is supposed that the contour ``$c$'' is monotonically 
increasing and regular, parameterized by a parameter $\tau\in{\Bbb R}$, 
$C=\{t\in{\Bbb C} \mid {\mbox Re}\,t = x_0(\tau), {\mbox Im}\,t = x_4(\tau), 
\tau \in {\Bbb R}\}$ and $\theta_c(t-t^\prime)=\theta(\tau-\tau^\prime)$. The 
spectral decomposition of $\widehat{\phi}$ in terms of plane waves has the ordinary 
form~\cite{LandsWeert}:
\begin{equation}
\widehat{\phi}(x)=\int\frac{d^3 {\bf k}}{(2\pi)^3 2\omega_k}
\left[a_k e^{-ikx}+a^{\dagger}_k e^{ikx} \right]\,\,,
\end{equation}
where $k_0=\omega_k=(|{\bf k}|^2+m^2)^{1/2}$. For compatibility with FTFT, we must
include the statistical distribution of the particles associated. For a single, scalar
field $\phi(x)$ we introduce the Bose-Einstein statistic given by
$N(k_0)=[e^{\beta k_0}-1]^{-1}$. In this case, combinations of creation and annihilation
operators are given by~\cite{LandsWeert}: 
\begin{align}
\langle a^{\dagger}_k a_k \rangle 
&=(2\pi)^3 2\omega_k N(\omega_k)\delta({\bf k}-{\bf k}^\prime)\nonumber \\[3mm]
\langle a_k a^{\dagger}_k \rangle
&=(2\pi)^3 2\omega_k \left[N(\omega_k)+1\right]\delta(\bf{k}-\bf{k}^\prime)\,\,,
\label{aa}
\end{align}
with the combinations of two creation or two annihilation operators vanishing.
The correlation functions $C^{>}(x,x^\prime)
=\langle\widehat{\phi}(x)\widehat{\phi}(x^\prime)\rangle=C^{<}(x^\prime,x)$ turns to 
have the following spectral expansion
\begin{equation*}
\langle\widehat{\phi}(x)\widehat{\phi}(x^\prime)\rangle = \int\frac{d^4 k}{(2\pi)^4}\,
e^{-ik(x-x^\prime)}\rho(k)\left[1+N(k_0)\right]\,\,,
\end{equation*}
where $\rho(k)=2\pi\left[\theta(k_0)-\theta(-k_0)\right]\delta(k^2-m^2)$. Their 
Fourier transforms, related by 
$\tilde{C}^{<}(k) = \rho(k)\left[N(k_0)+1\right]=e^{\beta k_0}\tilde{C}^{>}(k)$,
can be used in order to write the contour ordered propagator in the
form~\cite{LandsWeert}:
\begin{equation}
G^{c}(x,x^\prime)=\int\frac{d^4 k}{(2\pi)^4}\,
e^{-ik(x-x^\prime)}\rho(k)\left[\theta_c(t-t^\prime)+N(k_0)\right]\,\,.
\label{propag2}
\end{equation}
Another useful form is obtained after integration on $k_0$,
\begin{align}
G^{c}(x,x^\prime)
=&\,\,\theta_c(t-t^\prime)\int\frac{d^3{\bf k}}{(2\pi)^32\omega_k}
\left\{\left[N(\omega_k)+ 1\right]e^{-ik(x-x^\prime)}
+ N(\omega_k)e^{ik(x-x^\prime)}
\right\}\nonumber \\[3mm]
&+ \theta_c(t^\prime-t)\int\frac{d^3{\bf k}}{(2\pi)^32\omega_k} 
\left\{\left[N(\omega_k)+ 1\right]e^{ik(x-x^\prime)}
+ N(\omega_k)e^{-ik(x-x^\prime)}\right\}.
\label{propag3}
\end{align}
Although each possible contour would correspond to a specific formalism of FTFT,
there are restrictions on the contours due to the necessary analyticity of the 
correlation functions and the KMS condition \cite{LandsWeert}. These conditions
cause the support of the two point function to be analytic on the strip given
by $-\beta\leq{\mbox Im}\,(t-t^\prime)\leq\beta$, which on the closure the 
distributional character takes place. Furthermore, for the analyticity of 
$C^{>}(x,x^\prime)$, which because of the factor $\theta_c(t-t^\prime)$ has  
vanishing contributions to the propagator (\ref{propag1}) if $t^\prime$ 
succeeds $t$ on $C$, it is required that 
$-\beta\leq{\mbox Im}\,(t-t^\prime)\leq 0$. 
Conversely, for the analyticity of $C^{<}(x,x^\prime)$, with factor 
$\theta_c(t^\prime-t)$, it is required that 
$0\leq{\mbox Im}\,(t-t^\prime)\leq \beta$. Combining both relations one can 
conclude that if the complex time $t_1$ succeeds $t_2$ on $C$, then there 
follows that ${\mbox Im}\,t_2 \geq {\mbox Im}\,t_1$ which imposes that $C$ must
have a non increasing imaginary part. In other words, $C$ must have constant or
decreasing imaginary part. This is called the {\em monotonicity condition}.   

At this point, once the adopted approach does not depend on Feynman graphics 
calculations, we can proceed the analysis without the need in specializing to 
Minkowiskian RTF or Euclidean ITF parameterizations of the contour. From (\ref{propag3})
we select two typical distributions a temperature dependent piece, $G^{(c)\pm}_{\rm mat}$,
and a temperature independent piece, $G^{(c)\pm}_{\rm vac}$, whose labels refer to
{\em matter piece} and {\em vacuum piece}, respectively:
\begin{align*}
G^{(c)\pm}_{\rm mat}(x,x^\prime)
&=\int\frac{d^3{\bf k}}{(2\pi)^32\omega_k}
N(\omega_k)e^{\mp ik(x-x^\prime)}\,\,,
\\[3mm]
G^{(c)\pm}_{\rm vac}(x,x^\prime)
&=\int\frac{d^3{\bf k}}{(2\pi)^32\omega_k} 
e^{\mp ik(x-x^\prime)}\,\,.
\end{align*}
In terms of these distributions, the general contour propagator turns to be
\begin{align}
G^{c}(x,x^\prime)=&\,\,\theta_c(t-t^\prime)\left[ G^{(c)+}_{\rm mat}(x,x^\prime)+ 
G^{(c)+}_{\rm vac}(x,x^\prime) + G^{(c)-}_{\rm mat}(x,x^\prime)\right]
\nonumber \\[3mm]
&+ \theta_c(t^\prime-t)\left[ G^{(c)-}_{\rm mat}(x,x^\prime)+ 
G^{(c)-}_{\rm vac}(x,x^\prime) + G^{(c)+}_{\rm mat}(x,x^\prime)\right].
\label{propag4}
\end{align}


The structure of the propagators of FTFT suggests that, at a given order in 
perturbation series, the crossing products between matter and vacuum pieces 
would produce qualitatively different divergences.
Furthermore, one could expect it to have also a proliferation of divergent 
terms. Another possible distinction between FTFT and QFT version would be on the 
amount of arbitrariness through the contribution to the singular order besides
the establishment of a temperature dependent renormalization extension problem.
We are going to verify that in some sense the fact above does occur. The perception
of either of these points and their consequences would become
more difficult or not depending on the renormalization procedure adopted.

We now turn to investigate the divergent content of the distributions
$G^{(c)\pm}_{\rm mat(vac)}$ by calculating their WFSs.

\begin{theorem}
Only the temperature-independent part contributes to the ${WF}(G^c)$.
\end{theorem}

\begin{proof}
We proceed the prove using the stationary phase method (see for example~\cite{RS},
Section IX.10). The phases of the distributions above have all the same form $\mp ik(x-y)$.
It is useful to unify the notation as much as possible and represent them all by defining
the following integral
\begin{equation}
G^{(c)\pm}_{\rm mat(vac)}=\int\!\!\!\frac{d^3{\bf k}}{(2\pi)^3} 
\frac{\tilde{f}_{\rm mat(vac)}({\bf k};m^2,\beta)}{2\omega_k}\,\,
e^{\mp i\left[\omega(t-t^\prime)-{\bf k}\cdot({\bf x}-{\bf x^\prime })\right]}\,\,,
\label{pgen1}
\end{equation}
where $\tilde{f}_{\rm mat}({\bf k};m^2,\beta)=N(\omega_k)$ and
$\tilde{f}_{\rm vac}({\bf k};m^2,\beta)=1$. One can define the phase function $\varphi_{\pm}$,  
\begin{equation}
\varphi_{\pm}({\bf k}, x-x^\prime )
=\pm\left[(t-t^\prime)|{\bf k}|-({\bf x}-{\bf x^\prime })\cdot{\bf k}\right]\,,
\label{phasef}
\end{equation}
to obtain the following oscillatory integrals for the distributions:
\begin{equation}
G^{(c)\pm}_{\rm mat(vac)}=\int\!\!\!\frac{d^3{\bf k}}{(2\pi)^3} 
a_{\pm {\rm mat(vac)}}(t-t^\prime ,|{\bf k}|;m^2)
e^{-i\varphi_{\pm}({\bf k}, x-x^\prime )}
\,.
\label{pgen2}
\end{equation}
where
\begin{equation}
a_{\pm {\rm mat(vac)}}(t-t^\prime ,|{\bf k}|;m^2)
=\frac{\tilde{f}_{\rm mat(vac)}({\bf k};m^2,\beta)}{2\omega_k}
e^{\mp i\left[(\omega-|{\bf k}|)(t-t^\prime)\right]}
\end{equation}
is the asymptotic symbol. From the definition of the phase function 
(\ref{phasef}), one can easily see that 
it must be such that ${\mbox Im}\,(t-t^\prime)\leq 0$. Then, had the monotonicity
condition not previously selected the possible contours, the $\varphi_{\pm}$ would
be ill-defined. Both are in fact manifestations of the necessary analyticity of the
Green functions. The directions along which the phase in the integrand do not vary
satisfying $\partial_{\bf k}\varphi_{\pm}=0$ give us the following critical set,
\begin{align*}
{{\mathscr C}_\varphi}_{\pm}=&\Bigl\{(x-x^\prime 
=(0, {\bf 0}), k) \mid (k\neq 0) \in {\Bbb R}^4 \Bigr\} \\[3mm]
&\cup\Bigl\{(x-x^\prime, k) \mid ({\bf x}-{\bf x^\prime}\parallel {\bf k}\neq {\bf 0})
\in {\Bbb R}^3, (t-t^\prime)\in {\Bbb C},
{\bf k}\cdot({\bf x}-{\bf x^\prime})>0 \\[3mm]
& \mbox{\hspace*{15em}}
{\mbox Re}\,(t-t^\prime)=|{\bf x}-{\bf x^\prime}|,
{\mbox Im}\,(t-t^\prime)=0\Bigr\} \\[3mm]
&\cup\Bigl\{\left(x-x^\prime, k\right)
\mid ({\bf x}-{\bf x^\prime}\parallel {\bf k}\neq {\bf 0}) \in {\Bbb R}^3,
(t-t^\prime) \in {\Bbb C}, {\bf k}\cdot({\bf x}-{\bf x^\prime})<0, \\[3mm]
& \mbox{\hspace*{15em}}
{\mbox Re}\,(t-t^\prime)=-|{\bf x}-{\bf x^\prime }|,
{\mbox Im}\,(t-t^\prime)=0\Bigr\}\,\,.
\end{align*}
Though there is the restriction to those terms in (\ref{propag4}) which 
satisfy the monotonicity condition, from the additional condition 
${\mbox Im}\,(t-t^\prime)=0$, one can see that there are no contributions coming 
from the pieces of the contour with non-vanishing imaginary part. It has 
important consequences in the analysis of the WFS for the ITFs. 
Because the set of singular points of the WFS is a subset of 
${{\mathscr C}_\varphi}_{\pm}$, and the ITF-like pieces of the contour 
are such that ${\mbox Im}\,(t-t^\prime)>0$, one can conclude that the WFS associated
to ITF correlation functions are empty. The stationary phase manifold $\Lambda_\varphi$
is the set of points of the critical set having the non vanishing four momentum component
given by the gradients $\partial_\mu \varphi_{+}=(|{\bf k}|,-{\bf k})$ and
$\partial_\mu \varphi_{-}=(-|{\bf k}|,{\bf k})$. Then, 
\begin{align}
{\Lambda_\varphi}_{\pm}=&\Bigl\{(x-x^\prime 
=(0,{\bf 0}),(\pm\lambda|{\bf k}|,\mp\lambda{\bf k})) \mid {\bf k} \neq {\bf 0}
\in {\Bbb R}^3, \lambda \in {\Bbb R}_+ \Bigr\} 
\nonumber\\[3mm]
& \cup \Bigl\{(x-x^\prime, (\pm\lambda|{\bf k}|,\mp\lambda{\bf k}))
\mid ({\bf x}-{\bf x^\prime}\parallel {\bf k}\neq {\bf 0}) \in {\Bbb R}^3,
(t-t^\prime)\in\,{\Bbb C},\nonumber\\[3mm]
&\mbox{\hspace*{4em}}
\lambda \in {\Bbb R}_+, {\bf k}\cdot({\bf x}-{\bf x^\prime})>0,
{\mbox Re}\,(t-t^\prime)=|{\bf x}-{\bf x^\prime }|,
{\mbox Im}\,(t-t^\prime)=0 \Bigr\}\label{statph}\\[3mm] 
& \cup \Bigl\{(x-x^\prime, (\pm\lambda|{\bf k}|,\mp\lambda{\bf k}))
\mid ({\bf x}-{\bf x^\prime} \parallel {\bf k}\neq {\bf 0}) \in {\Bbb R}^3,
(t-t^\prime) \in {\Bbb C},\nonumber\\[3mm]
& \mbox{\hspace*{4em}}
\lambda \in {\Bbb R}_+, {\bf k}\cdot({\bf x}-{\bf x^\prime })<0,
{\mbox Re}\,(t-t^\prime)=-|{\bf x}-{\bf x^\prime}|, {\mbox Im}\,(t-t^\prime)=0
\Bigr\}\,\,.\nonumber
\end{align}
The result above can be interpreted as the set of pairs of which the critical
character of the phase is such that it breaks certain natural tendency  
of the integrals to converge due to its oscillatory character 
(see Riemann-Lebesgue Lemma~\cite{RS}). 
Such pairs are, therefore, suspect to be responsible for some bad behavior of 
the oscillatory integral. This implies that
${WF}(G^{(c)\pm})\subseteq\Lambda_{\varphi\pm}$ (again, see Section IX.10 in~\cite{RS}).
Because we are still able to save the convergence in some or even in all those critical
directions, there remains to be studied the contributions of the asymptotic symbols,
$a_{\pm {\rm mat(vac)}}$ and, in particular of $\tilde{f}_{\rm mat(vac)}$, to the convergence
of the integrals. 
For the temperature dependent part, to every possible contribution considered 
in the stationary phase manifold (\ref{statph}), the exponential factor 
$e^{\beta\omega_k}$ in the denominator of the integrand
$\tilde{f}_{\rm mat}({\bf k};m^2,\beta)=N(\omega_k)$ assures the condition for a 
fast decreasing function (see Sec. 2) to be fulfilled in every of those critical
directions. This guarantees the existence of the oscillatory integral and 
characterizes $G^{(c)\pm}_{\rm mat}$ to be a smooth function. Its WFS contribution 
is then empty. However, in the case of the vacuum piece, 
$\tilde{f}_{\rm vac}({\bf k};m^2,\beta)=1$, the factor $\frac{1}{\omega_k}$ does not
suffice to assure the asymptotic fast decrease in none of those critical 
directions. So, every pair in $\Lambda_{\varphi\pm}$ turns to be an element of 
the WFS. Therefore we have
\begin{eqnarray}
{WF}(G^{(c)\pm}_{\rm vac})=\Lambda_{\varphi\pm}\quad{\mbox{and}}\quad
{WF}(G^{(c)\pm}_{\rm mat})=\emptyset\,\,.
\label{wfsgs}
\end{eqnarray}
Hence, there are no contributions coming from the matter temperature-dependent part
to the ${WF}(G^c)$.
\end{proof}

It perhaps is necessary to emphasize that the $G^{(c)\pm}_{\rm vac}$ has
exactly the same singular spectrum as the Wightman function $\Delta_\pm$,
in Eq.(\ref{wfwight}), for of the ordinary QFT. Thus, we have settled that
\begin{equation}
{WF}(G^{(c)\pm}_{\rm vac})={WF}(\Delta_\pm).
\label{equalwfsgvdm}
\end{equation}
There follows then the same rules discussed for $\Delta_{\pm}$, in 
particular, for the product $\theta^\pm\cdot G^{(c)\pm}_{\rm vac}$ one has
\vspace*{-5mm}
\begin{align}
{WF}(\theta^\pm\cdot G^{(c)\pm}_{\rm vac})={WF}(\theta^\pm\cdot \Delta_\pm)
\quad{\mbox{and}}\quad
(x,0)\not\in({WF}(\theta^\pm)\oplus{WF}(G^{(c)\pm}_{\rm vac}))\,\,, 
\label{welldefg}
\end{align}
what characterizes it as well-defined and consequently, from the results of the
condition of the H\"ormander's criterion (\ref{HC}),
\begin{align}
{WF}(\theta^\pm\cdot G^{(c)\pm}_{\rm vac})\subset&\,\,{WF}(\theta^\pm)
\cup {WF}(G^{(c)\pm}_{\rm vac})\cup({WF}(\theta^\pm)\oplus{WF}(G^{(c)\pm}_{\rm vac}))
\nonumber \\[3mm]
=&\,\,{WF}(\theta^\pm)\cup{WF}(\Delta_\pm)
\cup ({WF}(\theta^\pm)\oplus{WF}(\Delta_\pm))\,.
\label{wfsgvth}
\end{align}
Because $G^{(c)\pm}_{\rm mat}$ is a smooth function from the Property 
\ref{wfsfgsmooth} in Remarks \ref{WFSPROP}, the product 
$\theta^\pm\cdot G^{(c)\pm}_{\rm mat}$ is such that
\begin{align}
{WF}(\theta^\pm\cdot G^{(c)\pm}_{\rm mat})\subset{WF}(\theta^\pm)
\quad{\mbox{and}}\quad (x,0)\not\in{WF}(\theta^\pm)\,\,.
\label{wfsgmth}
\end{align}
For this reason, in view of (\ref{welldefg}), (\ref{wfsgvth}) and (\ref{wfsgmth}),
the FTFT contour propagator $G^{(c)}$, (\ref{propag4}), is well-defined as sum of
well-defined products. From the Property \ref{wfssum} in Remarks~\ref{WFSPROP} 
and (\ref{superset}), we have that
\begin{align}
{WF}(G^{(c)})\subset&\,\,\Bigl[{WF}(\theta^+)\cup{WF}(\Delta_+)
\cup ({WF}(\theta^+)\oplus{WF}(\Delta_+)) \cup \nonumber \\[3mm]
\cup&\,\,{WF}(\theta^-)\cup{WF}(\Delta_-)
\cup ({WF}(\theta^-)\oplus{WF}(\Delta_-))\Bigr] \supset {WF}(\Delta_F)\,.
\label{equalwfs}
\end{align}
On the other hand, in the higher orders of the perturbation calculations there arise
products of propagators. In special, let us consider those terms in which there 
are products like $G^{(c)+}_{\rm vac} \cdot G^{(c)-}_{\rm vac}$. From (\ref{wfsgs}) one 
can see that in the same way of the ordinary QFT for $\Delta_+$,
\begin{align}
(x,0)\in{WF}(G^{(c)+}_{\rm vac}) \oplus {WF}(G^{(c)-}_{\rm vac})\,.
\label{vvill}
\end{align}
It does not match the condition for the H\"ormander's criterion. Indeed, this is
also an ill-defined product if the support of the distributions include
$x=0$, what turns it to be a problem to be treated through the renormalization 
procedure. But products like $G^{(c)s}_{\rm mat} \cdot G^{(c)s^\prime}_{\rm mat}$ and
$G^{(c)s}_{\rm mat} \cdot G^{(c)s^\prime}_{\rm vac}$, where $s,s^\prime=(+,-)$,
are well-defined because $G^{(c)s}_{\rm mat}$ are smooth functions. Therefore, by
considering products of propagators in the FTFT, both in RTF and ITF, one can expect
that the presence of the matter piece does not contribute to generate ill-defined
terms beside those already found in the ordinary QFT. Nevertheless, in the higher
orders in the perturbation expansion, it appears as temperature-dependent factors to
the ordinary divergences.  Roughly speaking, although the ill-defined products are the
same as the QFT ones, they appear with temperature-dependent factors. 

Another aspect of the renormalization concerns the arbitrariness or ambiguity of
the process and its relation to physical symmetries. The amount of arbitrariness is
governed by the singular order and scaling degree of the distributions
involved~\cite{Ste1}-\cite{BF}. Once for $G^{(c)\pm}_{\rm mat}$, 
\begin{align*}
(G^{(c)\pm}_{\rm mat})_\lambda 
&= G^{(c)\pm}_{\rm mat}(\lambda(x-x^\prime);m^2,\beta)\!
=\!\int\frac{d^{(d-1)}{\bf k^\prime}}{(2\pi)^3\,2\omega_{k^\prime}} 
N(\omega_{k^\prime}) e^{\mp ik^\prime\lambda(x-x^\prime)} \\[3mm]
& = \lambda^{2-d}G^{(c)\pm}_{\rm mat}(x-x^\prime;\lambda^2m^2,\lambda^{-1}\beta)\,\,,
\end{align*}
then, one has $\omega\geq d-2$, the scaling degree is $\sigma(G^{(c)\pm}_{\rm mat})=d-2$ 
and singular order is $\Sigma(G^{(c)\pm}_{\rm mat})=-2$. Notice further that 
\begin{equation}
\Sigma(G^{c\pm}_{\rm mat})=\Sigma(\Delta_\pm)=\Sigma(G^{(c)\pm}_{\rm vac})\,\,.
\label{correlsd}
\end{equation} 
Hence, for the FTFT propagator $G^{(c)}$ in (\ref{propag4}), we obtain that 
\begin{align}
\sigma(\theta^\pm G^{c\pm}_{\rm mat(vac)})&=\sigma(G^{(c)})=\sigma(\Delta_F)=d-2\,,
\label{propgsd}\\[3mm]
\Sigma(\theta^\pm G^{c\pm}_{\rm mat(vac)})&=\Sigma(G^{(c)})=\Sigma(\Delta_F)=-2\,.
\label{propgso}
\end{align} 

We shall analyze, as a representative case of the higher order product in 
the perturbation series, the square of the propagator associated to the branch 
of the contour which is parameterized forward in the real time only. We consider
again the products of propagators arising in the perturbation series. 
For the products like $G^{(c)s}_{\rm mat} \cdot G^{(c)s^\prime}_{\rm mat}$, 
$G^{(c)s}_{\rm mat} \cdot G^{(c)s^\prime}_{\rm vac}$ and 
$G^{(c)s}_{\rm vac} \cdot G^{(c)s^\prime}_{\rm vac}$ we have 
\begin{align}
\sigma(G^{(c)s}_{\rm mat(vac)} \cdot G^{(c)s^\prime}_{\rm mat(vac)})&=2(d-2)
\nonumber\\[3mm]
\Sigma(G^{(c)s}_{\rm mat(vac)} \cdot G^{(c)s^\prime}_{\rm mat(vac)})
&=2(d-2)-d\,\,.  
\label{sigmaprods}
\end{align} 
The scaling degree and the singular order are the same for both 
the matter or vacuum pieces products. In view of this,
one can see that the singular order determines the number of arbitrary coefficients 
(counter terms) in the renormalization procedure. As a simple example, let us examine
an 1-loop diagram in $\frac{g}{4!}\phi^4$, a truncated $4$-point diagram
with two internal lines connecting two different vertex,
\begin{align}
\Gamma^{(4)}\sim g^2 [G^{(c)}(x-x^\prime)]^2 = g^2 
&\left\{\sum_{s=+,-}\theta^s \theta^s G^{(c)s}_{\rm vac} G^{(c)s}_{\rm vac} \right.  
+\sum_{s=+,-}\theta^s \theta^{-s} G^{(c)s}_{\rm vac} G^{(c)-s}_{\rm vac} + 
\nonumber \\[3mm]
&+2\sum_{s\,\,s^\prime}\theta^s \theta^s G^{(c)s}_{\rm vac} G^{(c)s^\prime}_{\rm mat} 
+2\sum_{s\,\,s^\prime}\theta^s \theta^{-s} G^{(c)s}_{\rm vac} G^{(c)s^\prime}_{\rm mat} 
+ \label{twopoints} \\[2mm]
&+\sum_{s\,\,s^\prime s^{\prime\prime}}
\theta^s \theta^s G^{(c)s^\prime}_{\rm mat} G^{(c)s^{\prime\prime}}_{\rm mat} 
+ \left. \sum_{s\,\,s^\prime \,\,s^{\prime\prime}}
\theta^s \theta^{-s} G^{(c)s^\prime}_{\rm mat} G^{(c)s^{\prime\prime}}_{\rm mat} 
\right\}\,\,.\nonumber
\end{align}     
Notice that the sum of products of distributions falls into different categories. As it 
was shown from (\ref{wfsgs}) to (\ref{vvill}) and in the chain of reasoning just after,
the only term which exhibits an ill-defined product is the second one. That product is
well-defined elsewhere, except at $x-x^\prime=0$. This is the target of the renormalization
in the present case. The degree of arbitrariness is governed by (\ref{sigmaprods}) and the
number of counter terms is limited by certain physical symmetries.

Next, we consider an overlapping higher loop with three internal lines connecting
two vertices,\footnote{Others Feynman diagrams can be composed by convolutions of propagators.
In essence, the presence of convolutions contribute to the well behavior of the
product of distributions, by decreasing the singular order and 
improving the fast decay of the symbols.}
\[
\Gamma^{(2)}\sim g^2 [G^{(c)}(x-x^\prime)]^3\,\,.
\]
One finds an abundance of ill-defined terms as compared to the ordinary QFT case. There will
appear ill-defined products like
\[
\theta^s \theta^s \theta^{-s} G^{(c)s}_{\rm vac}G^{(c)s}_{\rm vac} G^{(c)-s}_{\rm vac}
\quad {\mbox{and}} \quad
\theta^s \theta^s \theta^{-s} G^{(c)s}_{\rm mat}G^{(c)s}_{\rm vac} G^{(c)-s}_{\rm vac}\,\,.
\]
The former suffers from the same illness as the second
term of (\ref{twopoints}), though it has a different degree and it is also to be treated
in a temperature-independent fashion. The latter, due to the presence of a matter piece
factor, in view of (\ref{sigmaprods}), could indicate a temperature-dependent renormalization
problem. However, from (\ref{wfsgs}) and (\ref{vvill}), one can easily see verify that such
an ill-definition is due to the product of vacuum pieces only. Notice that this term 
was treated in a temperature-independent way in the lower order term (\ref{twopoints})
and that the temperature-dependent part appears as simple factor. For the general case of
the doubling of degrees of freedom with contour parameterized both forward and backward
in the real time, the analysis is similar. 

This quantitative analysis has shown that, despite the existence of temperature-dependent
factors multiplying the ill-defined products, from the point of view of the renormalization
problem, it can be treated order by order as a vacuum renormalization problem. Furthermore,
the degree of arbitrariness in the process for a given order is limited by the singular
order of the temperature-independent piece and, from products of them, there arise at each
order an ill-defined product that is leading in singular order and degree of arbitrariness.
This makes clear that the matter piece, being absent from a singular spectrum, cannot
include any new contribution to FTFTs concerning the category of ill-defined products yet
found in the ordinary QFT.

\section{Conclusions}
The results concerning the renormalization of FTFT has been extensively analyzed
in the literature. The present contribution lies on the method which allows us to clarify
some points in the comparison between QFT and FTFT renormalization.
The problem of the divergences was faced from the ground by
the mathematical study of the basic ill-defined products distributions, {\em i.e.},
the lack of definition of the distributional product on the coinciding points.
An important role was played by the microlocal analysis. By using the H\"ormander's criterion,
based on the WFSs of distributions, we have shown that the contribution to form ill-defined
products comes from the temperature-independent pieces only. Hence, the matter piece does
not contribute to form divergent terms though it can appear as factors of the divergent ones. 
The structure of the propagators of FTFT, being separable into vacuum and matter pieces
turns easy the analysis of the ill-defined products. As a matter of fact, one shows that
the separation also generates an increasing on the number of ill-defined products in the
perturbation series due to the mixing of these factors in crossing products in the higher
order terms of the perturbation expansion. The matter piece appears then as
temperature-dependent factors of ill-defined products vacuum pieces in the higher orders of 
the perturbation series. Focusing on the perturbation series, the degree of arbitrariness
in the process for a given order is determined by the temperature-independent ill-defined 
product leading in singular order. Hence, the problem of the extension reduces 
at each order to the analogous one of the ordinary QFT. Consequently, it is 
proved that the amount of arbitrariness in the renormalization procedure, 
as well as the type of the ambiguities, if conveniently treated, remains the same 
when passing from a given QFT to the associated FTFT version. The perception of 
either of these points and their consequences could be difficult or not 
depending on the renormalization procedure adopted.

Applications of the results given in this paper will appear in a coming
paper~\cite{DanZe}, where we study the renormalization of the eletromagnetic and
gravitacional couplings of an electron which is immersed in a heat bath under
the light of the scheme of Brunetti-Fredenhagen-Holland-Wald~\cite{BF,HW}, who
have demonstrated renormalizability of QFTs satisfying the requirements of
Weinberg's theorem on general curved space-times using a microlocal adaptation
of the Epstein-Glaser approach. Our aim is to clarify the connection between microlocal
analysis and the area of QFT at a finite temperature.

\section*{Acknowledgements}
J.L. Acebal was supported by the Brazilian  agency Conselho Nacional de Desenvolvimento
Cient\'\i fico e Tecnol\'ogico (CNPq).


\end{document}